\documentclass[openacc]{rstransa}

\usepackage[version=4]{mhchem}
\usepackage{float}
\usepackage{natbib}
\usepackage{subcaption}

     \def\aap{A\&A}
     \def\apjl{ApJL}
     \def\apjs{ApJS}

\begin{document}

\title{Lightning and charge processes in brown dwarf and exoplanet atmospheres}

\author{
Christiane Helling$^{1,2}$ and Paul B. Rimmer$^{3,4,5}$}

\address{$^{1}$Centre for Exoplanet Science, School of Physics \& Astronomy, University of St Andrews, North Haugh, St Andrews, KY16 9SS\\

$^2$ SRON Netherlands Institute for Space Research, Sorbonnelaan 2, 3584 CA Utrecht, NL\\
$^{3}$Department of Earth Sciences, University of Cambridge, Downing St, Cambridge CB2 3EQ\\
$^{4}$Cavendish Astrophysics, JJ Thomson Ave, Cambridge CB3 0HE\\
$^{5}$MRC Laboratory of Molecular Biology, Francis Crick Ave, Cambridge CB2 0QH}

\subject{xxxxx, xxxxx, xxxx}


\corres{Christiane Helling\\
\email{ch80@st-andrews.ac.uk}}

\begin{abstract}
The study of the composition of brown dwarf atmospheres helped to understand their formation and evolution. Similarly, the study of exoplanet atmospheres is expected to constrain their formation and evolutionary states. We use results from 3D simulations, kinetic cloud formation and kinetic ion-neutral chemistry to investigate ionisation processes which will affect their atmosphere chemistry: The dayside of super-hot Jupiters is dominated by atomic hydrogen, and not H$_2$O. Such planetary atmospheres exhibit a substantial degree of thermal ionisation and clouds only form on the nightside where lightning leaves chemical tracers (e.g. HCN) for possibly long enough to be detectable. External radiation may cause exoplanets to be enshrouded in a shell of highly ionised, H$_3^+$-forming gas and a weather-driven aurora may emerge. Brown dwarfs enable us to study the role of electron beams for the emergence of an extrasolar, weather-system driven aurora-like chemistry, and the effect of strong magnetic fields on cold atmospheric gases. Electron beams trigger the formation of H$_3^+$ in the upper atmosphere of a brown dwarf (e.g. LSR-J1835) which may react with it to form hydronium, H$_3$O$^+$, as a longer lived chemical tracer. Brown dwarfs and super-hot gas giants may be excellent candidates to search for H$_3$O$^+$ as an H$_3^+$ product.
\end{abstract}



\maketitle
\section{Introduction}

Exoplanet science is moving from object discoveries (e.g. \citealt{2014PNAS..11112647B}) into characterisation and analysis of the discovered objects (e.g. \citealt{2007Natur.448..169T,2009ApJ...699..478D,2018AJ....156...17K,2018Natur.557..526N,2018A&A...615A..16B}). Transit spectroscopy has shown that exoplanet atmospheres form clouds (e.g. \citealt{2011MNRAS.416.1443S}), and that phase curves of exoplanets indicate the presence of winds driven by the external, host-star irradiation (e.g. \citealt{2015ApJ...814L..24L,2007Natur.447..183K}). The formation of extrasolar  clouds and their time-variability has first been analysed in brown dwarf atmospheres \citep{2002ApJ...577..433G,2012ApJ...750..105R,2013ApJ...768..121A} which are repeatedly studied as exoplanet analogs \citep{2019MNRAS.483..480V}. Complex models have been developed for brown dwarf and planetary atmospheres, their chemical composition and the clouds that form \citep{2014MNRAS.445..930C,2012ApJ...754..135M,2016A&A...594A..48L,2017ApJ...836...73Z,2018A&A...615A..97L}. 

In fact, clouds have been found to be quiet frustrating as they are blocking our view into the extrasolar  and chemically very different atmospheres of exoplanets  where one might hope to find the signature of biomolecules or the precursors thereof. Because the atmospheric chemistry of exoplanets (and brown dwarfs) is very different to Earth, the clouds that form  are not made of water only, but are made of a mix of minerals and oxides in the hotter exoplanets in addition to water or methane clouds in cooler exoplanets and brown dwarfs \citep{2018arXiv181203793H}. Although the Solar System gas planet atmospheres appear less dynamic compared to some of the extrasolar gas giants; lightning events are detected directly in the cloudy atmospheres of Earth, Jupiter, and Saturn, are debatable for Venus, and indirectly inferred for Neptune and Uranus. Sprites and high-energy particles are observed above thunderclouds on Earth \citep{2013SGeo...34....1F,2013ERL.....8c5027F}, and are  predicted for Jupiter, Saturn, and Venus. Lightning observations can only be conducted in situ on Earth such that the lightning statistics for all other Solar System planets are rather incomplete \citep{2016MNRAS.461.3927H}. Possible analogs for exoplanet lightning or lightning in brown dwarfs are terrestrial volcanoes that produce lightning during an eruption in their plumes. Studying lightning in other planets inside and outside our Solar System is of interest because it enables us to study the electrostatic character of such alien atmospheres, and it opens new possibilities for tracking the dynamic behaviour and the associated chemical changes in such extraterrestrial environments. Other astronomical objects, brown dwarfs and planet-forming disks, are also expected to have lightning discharges \citep{2016PPCF...58g4003H,2000Icar..143...87D} because the underlying physical regimes are similar  \citep{2016SGeo...37..705H}. 

Charge processes on exoplanets and brown dwarfs are furthermore driven by their environments. The global environment is very different for exoplanets and for brown dwarfs, reflecting their different formation mechanisms.  The exoplanet's atmosphere is exposed to the radiation field of its host star and the effect of it will also depend on the exoplanet's size, mass, atmosphere thickness, and its distance from the host star. A seemingly small difference in orbital distance and planetary mass leads to vastly different atmospheric conditions, deciding between habitable (Earth), poisoning (Venus) or too extreme (Mars) for life as we know it. A brown dwarf is exposed to the insterstellar radiation field, or when being part of a binary system with a white dwarf, it may suffer the harsh radiation from the nearby white dwarf  \citep{2017MNRAS.471.1728L,2018MNRAS.481.5216C}. In contrast to planets, brown dwarfs are  magnetically very active with B\,$\approx 10^3$G  \citep{2002ApJ...572..503B}, and a current system may cause electrons to collide with the atmospheric hydrogen. The highly polarized kHz and MHz radio emission has been interpreted as Auroral emission, comparable to Jupiter but $10^4\times$ stronger \citep{2015Natur.523..568H}.

Predictions of significant \ce{H_3^+} concentrations in Jupiter's upper atmosphere \citep{Gross1964, Atreya1976}, lead to observational programme that successfully detected \ce{H_3^+} in Jupiter, {\it in situ} \citep{Hamilton1980}, and remotely \citep{Drossart1989}. Likewise, the ionisation of exoplanet atmospheres is predicted to lead to the formation of upper atmospheric \ce{H_3^+} \citep{Miller2000}, and this prediction has lead to several observational programs to search for \ce{H_3^+} in Hot Jupiter atmospheres, thus far without success \citep{Goto2005,Shkolnik2006,2016A&A...589A..99L}. Although there are spectral hints of \ce{H_3^+} in brown dwarf atmospheres \citep{2015MNRAS.447.3218C}, no definitive detection of \ce{H_3^+} has been made in a brown dwarf atmosphere, and there have been as yet no published results from a program to search for \ce{H_3^+} in brown dwarf atmospheres. We will allude to why it may be difficult to detect \ce{H_3^+} on the targeted exoplanets and brown dwarfs in this paper.

In what follows, we discuss environmental processes that affect the ionisation of exoplanet and brown dwarf atmospheres that cause the formation a global or partial ionosphere (Sect.~\ref{s:GI}). The term 'aurora' is utilised for a range of processes involving accelerated electrons in the upper atmosphere of the Solar System planets. Here, {\it an aurora is understood to be associated with a pool of free charges, a confining field and collisional partners for accelerated charges.} An ionosphere therefore provides one of the necessary condition for an aurora to emerge on brown dwarfs and exoplanets. The conditions for lightning that locally changes the thermodynamic conditions dramatically, and that may leave some tracer molecules, are laid out for exoplanets and brown dwarfs in Sect.~\ref{s:GI}\ref{s:LI}. We proceed to introduce how extrasolar aurorae might be traced by chemical signatures beyond H$_3^+$, namely by hydronium,  H$_3$O$^+$ in Sect.~\ref{s:ALSR}. We suggest brown dwarfs as optimal candidates to search for H$_3$O$^+$ as a product of H$_3^+$.

\begin{figure}[ht]
{\ }\\*[-3.5cm]\centering
\includegraphics[width=0.9\textwidth]{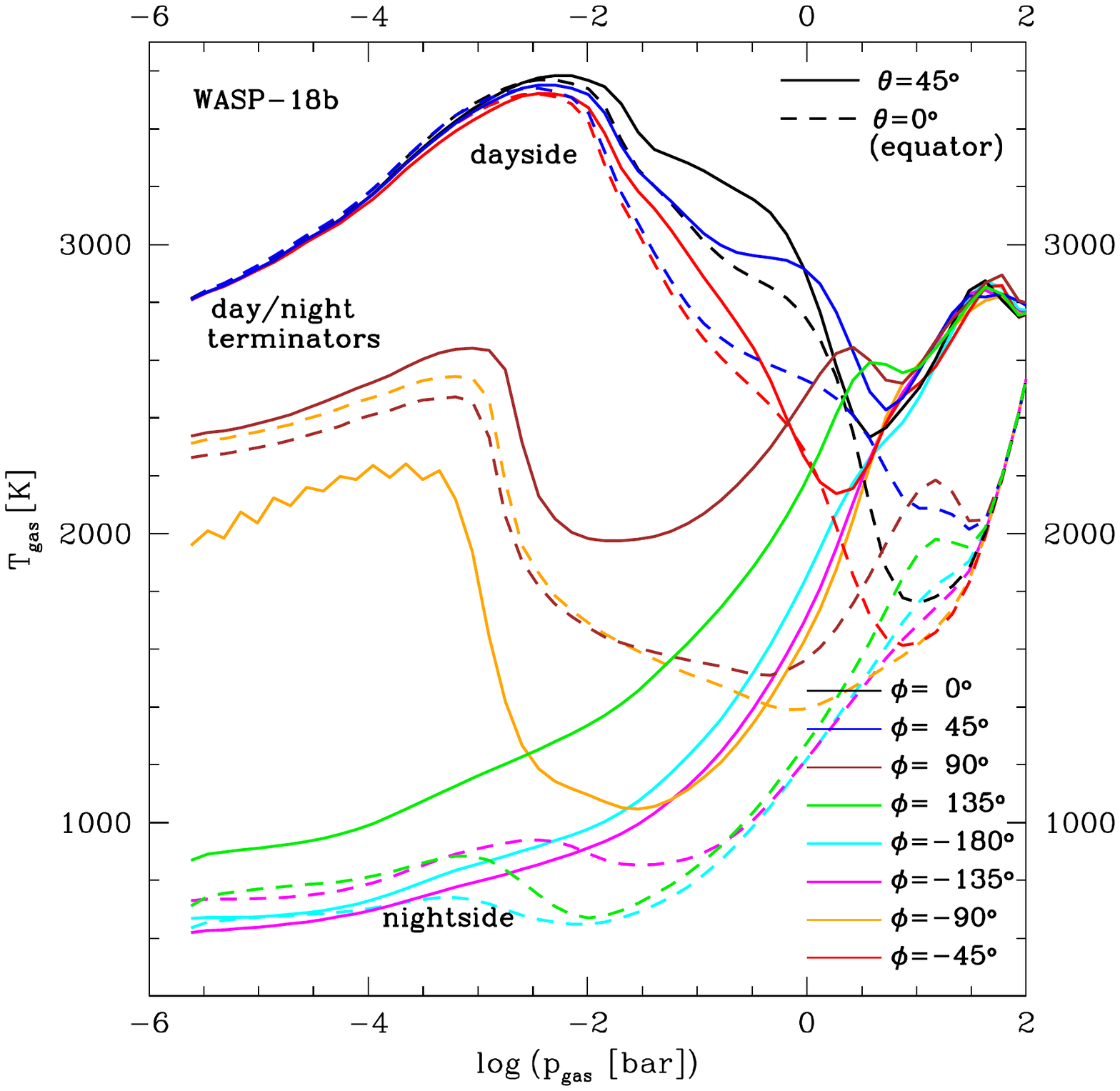}\\*[-4.0cm]
\caption{Highly irradiated, ultra-hot Jupiters develop extreme temperature differences of 2500K between day- ($\phi=-45^o, 0^o, 45^o$) and nightside ($\phi=135^o, -180^o, -135^o$). An ionosphere may emerges on the dayside, and mineral clouds form on the nightside \citep{2019arXiv190108640H}. The terminator regions ($\phi=90^o, -90^o$) are transition regions between the two extreme atmosphere conditions. The 1D profiles are from a cloud-free 3D GCM simulation \citep{2018A&A...617A.110P}.}
\label{figTpWASP18b}
\end{figure}

\section{Ionisation processes on exoplanets and brown dwarfs}\label{s:GI}

The best studied exoplanets to date are the short-period giant gas planets  HD\,189733b and HD\,209458b,  and super-hot gas giants like WASP-18b, WASP-121b or HAT-P-7b, see \citep{2016MNRAS.460..855H,2019arXiv190108640H,2018A&A...615A..97L,2018A&A...617A.110P,2015A&A...580A..12L} and references therein. These exoplanets, despite being clearly uninhabitable, enable us to test our models and expand our ideas into the extreme regimes of exoplanetary atmospheres. For example, molecules as tracers for atmospheric processes have long been studied in cool stars (see \cite{1997IAUS..178..441J} for a review), and the respective model atmosphere expertise is the backbone of exoplanet and brown dwarf atmosphere research \citep{1996A&A...308L..29T,1997ARA&A..35..137A}.  Low-gravity brown dwarfs turn out to be rather suitable analogues for long-period extrasolar giant gas planets and almost overlap in the colour-magnitude diagram (see Fig. 1 in \citealt{2018ApJ...854..172C}). White dwarf - brown dwarf binaries can be seen as analogues to short-period giant gas planets \citep{2018MNRAS.481.5216C}.

The ionisation processes in brown dwarf and exoplanet atmospheres are determined by their individual environments and more locally by the objects' character (size, mass, age) itself. The dominating ionisation processes in exoplanets and brown dwarfs are thermal ionisation (Brownian motion) of the gas, tribolelectric charging (turbulent motion) of cloud articles, and photoionisation which can affect both, gas and cloud particles. Lightning and Alfv{\'e}n ionisation produce short-lived population of charges in the atmosphere.

\begin{figure}[!h]
\hspace*{-1.4cm}\includegraphics[width=1.2\textwidth]{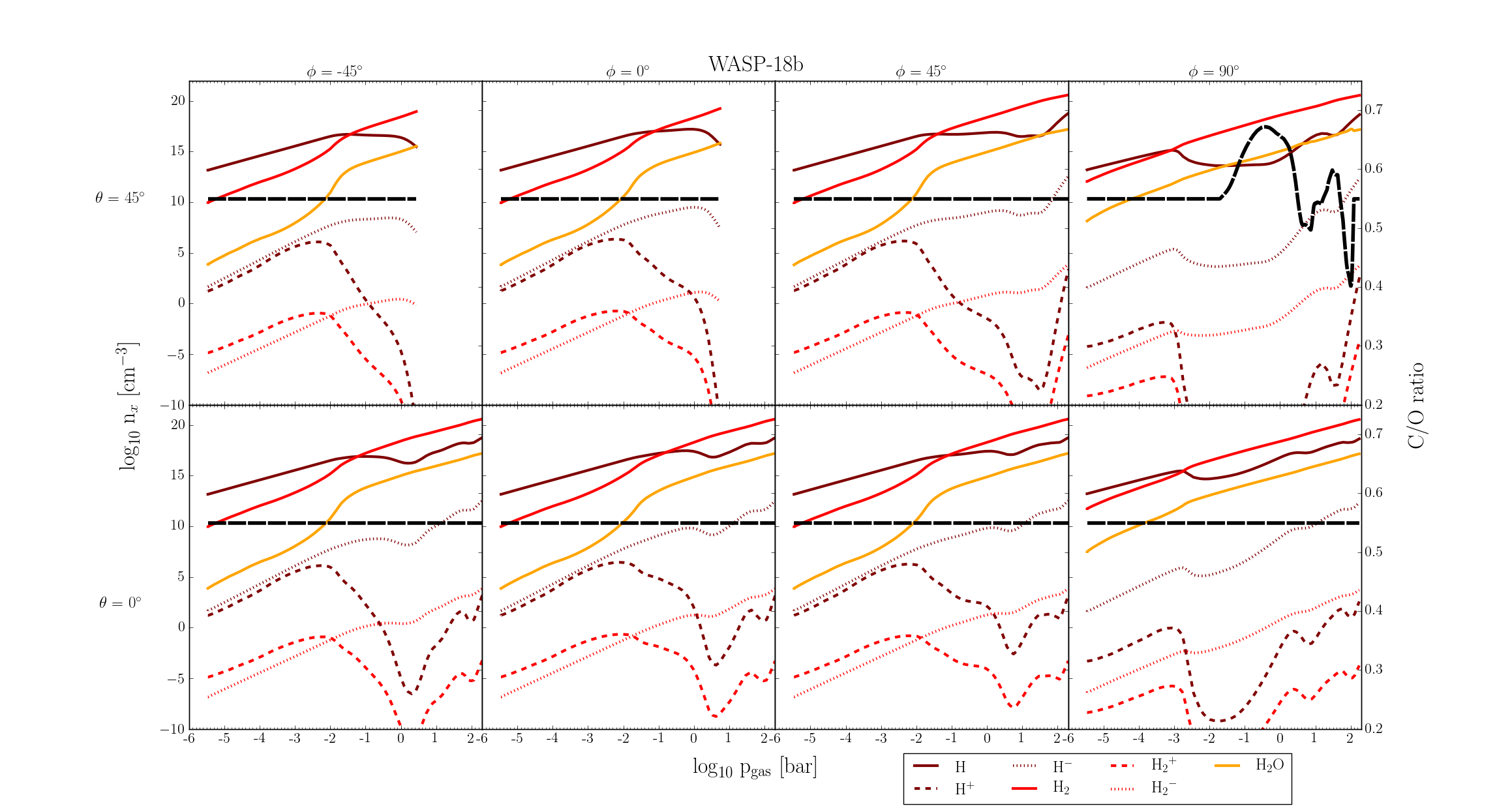}\\
\hspace*{-1.4cm}\includegraphics[width=1.2\textwidth]{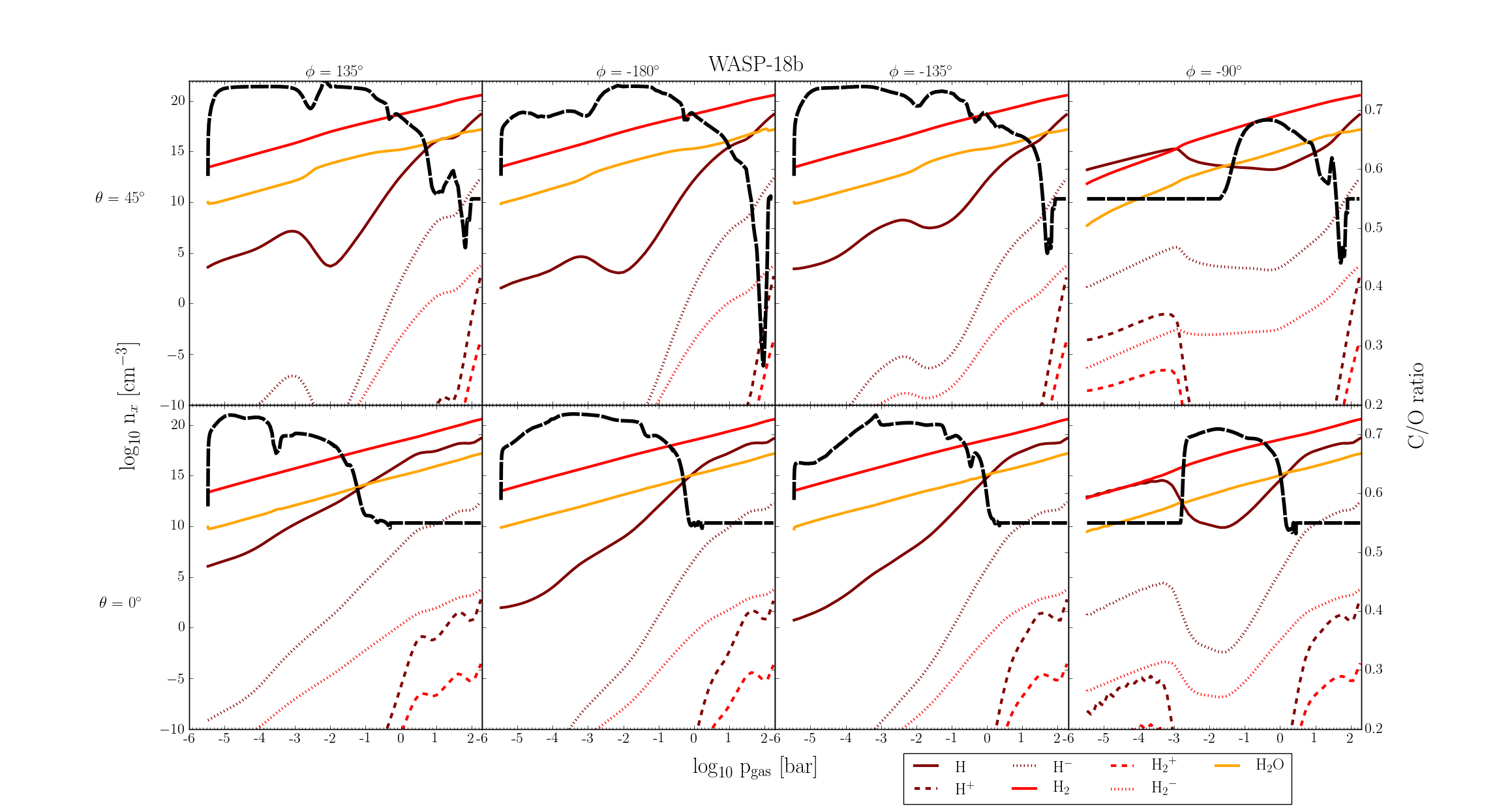}
\caption{The changing number densities ($n_{\rm x}$ [cm$^{-3}$], left axis) of hydrogen-binding gas species from the dayside (top) to the nightside (bottom). C/O is shown as black-dashed line (right axis) to demonstrate where cloud formation takes place. The dayside is made of a cloud-free,  H-dominated gas and the night-side is made of a H$_2$-dominated gas with vivid cloud formation (C/O>C/O$_{\rm solar}$). The extension of the cloud in pressure space (x-axis) changes between equator ($\theta=0$) and norther hemisphere ($\theta=45^o$);  \citep{2018A&A...617A.110P,2019arXiv190108640H}.}.
\label{figHWASP18b}
\end{figure}

\subsection{Thermal ionisation and the day/night differences on ultra-hot Jupiters}\label{ss:WASP121b}
The high irradiation that close-in planets with a rather small orbital distance to their host star receive can lead to a drastic temperature difference between day and night side. That results in large pressure gradients which drive a strong global circulation. For example, the 3D global circulation model for WASP-18b from \citep{2018A&A...617A.110P} suggests $T_{\rm day}-T_{\rm night}\approx 2500$K (Fig.~\ref{figTpWASP18b}), but global circulation can also be driven by  more moderate day-night-differences \citep{2013MNRAS.435.3159D,2018arXiv180708433S}. As show in  Fig.~\ref{figTpWASP18b} for the super-hot Jupiter WASP-18b, the vertical dayside temperature profiles ($\phi=-45^o, 0^o, 45^o$)  have strong temperature inversions, i.e. outward increasing gas temperatures, which become more shallow in the terminator regions ($\phi=90^o, -90^o$)  where they occur at lower pressures compared to the dayside. The nightside temperature ($\phi=135^o, -180^o, -135^o$) smoothly decreases outwards, and are very similar to those of non-irradiated brown dwarfs and planets that orbit their host star at a large distance (so-called 'directly imaged' planets). The effect of the high irradiation appears smaller for higher latitudes ($\theta=45^o$, dashed lines in Fig.~\ref{figTpWASP18b}). The depth of the temperature inversions is less at higher latitudes and the day-night temperature differences is smaller than in the equatorial regions.

The extreme day-night temperature differences produce very different chemical structures of the day- and the night-side, and transition regions at the terminators (Fig.~\ref{figHWASP18b}). On the dayside (top), all H$_2$ is thermally dissociated and the atmosphere is dominated by the atomic hydrogen (H). No H$_2$O can form because of the high temperatures. Atoms like Na, K, Mg, but also Al and Ti undergo thermal ionisation causing the local degree of thermal ionisation to change by 12 order of magnitudes from the night- to a dayside value of $10^{-3.5}$.  The plasma frequency, $\omega_{\rm pe}\sim(n_{\rm e}/m_{\rm e})^{1/2}$, is $10^4\times$ the electron-neutral collision frequency, $\nu_{\rm ne}\sim n_{\rm gas} v_{\rm them, e}$, hence, the electromagnetic interactions dominate over kinetic collisions between the electrons and neutrals. A current system similar to that in the Earth's atmosphere may be expected and a weather-driven aurora may emerge if the planet possesses a confining field. On the cold night-side (bottom), H$_2$ remains the dominating gas species with also no other species being thermally ionised in the upper, cool part of the atmosphere. The next most abundant H-binding molecule after H$_2$ is H$_2$O. Cloud formation causes a depletion of oxygen which causes the carbon-to-oxygen ration, C/O, to increase to > 0.7 \citep{2019arXiv190108640H}. Figure~\ref{figHWASP18b} depicts C/O as tracer for the cloud location in the atmosphere (black dashed line). Lightning activity can be expected on the cloud-forming night side of super-hot gas giants and that the photoionisation of H$_2$ may produce H$_3^+$ because of a very low degree of ionisation and a slow dissociative recombination of H$_3^+$ (see Sect.~\ref{s:ALSR}\ref{ss:h3+}).

\subsection{H$_3^+$ in an envelope of highly ionised gas on brown dwarfs?}\label{ss:environ}

The radiative environment of exoplanets and brown dwarfs affect the atmospheric chemistry due photon-chemistry processes but also due to thermal processes. Exoplanets are always exposed to the radiation field of their host stars (the flux of which scales with $r^{-2}$, $r$ being the star--planet distance), and it will be the harsh interstellar radiation field or even the radiation from a white dwarf that affects a brown dwarf's atmosphere \citep{2015MNRAS.447.3218C}. The white dwarfs radiation field dissociates H$_2$ in the upper atmosphere of brown dwarfs with observed emission from H$\alpha$, He, Na, Mg, Si, K, Ca, Ti and Fe \citep{2017MNRAS.471.1728L}. Photo-chemical processes triggered by Lyman continuum radiation further dissociates atomic hydrogen. A completely ionised outer atmosphere environment results for brown dwarfs as white dwarf companions, and to a lesser degree for the interstellar radiation field or the Lyman continuum flux from high mass O/B stars \citep{2018A&A...618A.107R}.  We note that the photoionisation of H$_2$ can lead on to the formation of H$_3^+$ which may be observable if its destruction due to dissociate recombination (Reaction~\ref{eq:direc} in  Sect.~\ref{s:ALSR}\ref{ss:h3+}) is inefficient.  This occurs if the Reactions~\ref{eqn:ionize} and~\ref{eqn:ionize2} dominate over Reaction~\ref{eq:direc} or if the electrons are removed quickly enough, so that Reaction~\ref{eq:direc} can not occur. \cite{2018A&A...618A.107R} have shown that the high atmosphere regions, where Lyman continuum radiation is effective, already contain a small fraction of small cloud particles. Electron attachment onto these cloud particles might decrease the efficiency of Reaction~\ref{eq:direc} such that H$_3^+$ remains in the gas phase. The same may be achieved by magnetically coupling an electron to a brown dwarf's strong magnetic field of 1000--6000G \citep{2018ApJS..237...25K}. The magnetic field strength required to assure magnetic coupling of an electron is mainly dependent on the density of the collisional partners, $n_{\rm gas}$, and the electron temperature, $T_{\rm e}$, such that   $B_{\rm e}\sim n_{\rm gas} (m_{\rm e}T_{\rm e})^{1/2}$ \citep{2015MNRAS.454.3977R}. Hence, this threshold decreases with the decreasing gas density in the upper atmosphere where H$_3^+$ forms as the result of photoionisation of H$_2$. Such localised interaction of an ionospheric environment with a local magnetic field has been traced through H$_3^+$ emission on Jupiter \citep{2018NatAs...2..773S}. \cite{2016A&A...589A..99L} reported a non-detection of H$_3^+$ for the hot-Jupiter HD\,209458b in secondary eclipse for spectroscopic observation of the planet's day-side. In the light of the above discussion, this implies that i) the magnetic field is too weak to enable a sufficient electron acceleration to enable H$_2$ dissociation on HD\,209458b,  ii) that the local supply of electrons is too low for enabling enough collisions with H$_2$ to occur (e.g. due to atmosphere being shielded by the host star from interstellar radiation field, atmosphere has not developed an ionosphere and remains at its hottest LTE dayside temperature of 1800K (see  \cite{2015A&A...580A..12L,2018A&A...615A..97L}), iii) a high H$_2$O or CO abundance destroys H$_3^+$ kinetically. We address the possibility of H$_3^+$ formation in brown dwarfs with our kinetic modelling approach in Sect.~\ref{s:ALSR}.

\subsection{Lightning on exoplanets and brown dwarfs}\label{s:LI}

Brown dwarfs and many extrasolar planets will form clouds in their atmospheres. The cloud particles charge due to photochemical processes nearer the top of the cloud (similar to what was described in Sect.~\ref{s:GI}\ref{ss:environ}) and by tribolectric processes due to turbulence driven particle-particle collisions \citep{2011ApJ...737...38H}. The charge that a cloud particle acquires increases with the size of the particle. As cloud particles gravitationally settle into deeper layers of the atmospheres and bigger particles may fall faster, a large-scale charge separation becomes established  and an electrostatic potential difference can build up inside a cloud. The stored energy may become large enough to overcome the local break-down potential such that a large-scale lightning inside these extraterestrial clouds emerges. The lightning discharge converts an initially semi-neutral gas of a moderate temperature of $\sim 1000 - 2000$K into a plasma channel of $\sim 30000$K which sends a shock wave into its immediate surrounding. While the discharge process will be relatively short-lived (of the order of seconds), the effect it has on the gas chemistry can prevail for longer.  

\citep{2016MNRAS.461.1222H,2017pre8.conf..345H} have investigated how much lightning would be required to reproduce a transient, unpolarised one-off radio signal of the exoplanet HAT-P-11b \citep{2013A&A...552A..65L}. The parameters for one lightning strike were adopted from Saturn and from Earth. The caveats with these assumptions are that HAT-P-11b is a mini-Neptune with $M_{\rm P}=26\,M_{\rm Earth}$ and $R_{\rm P}=4.7\,R_{\rm Earth}$ \citep{2016MNRAS.461.3927H}, and all lightning measurements are carried out for Earth and detection for Solar System planets only. Using lightning strike statistics is therefore a formidable task, given that the Jupiter and Saturn measurements must be considered as incomplete (see discussion in \cite{2016MNRAS.461.1222H}). While it is reasonable to assume that the electrostatic breakdown field does not vary strongly with the local composition of the gas (as demonstrated in \cite{2013ApJ...767..136H}), the radiation power derived for HAT-P-11b at a distance of d=38pc from Earth for one lightning flash of $2.2\,10^{14}$ Jy is based on values for  a  Saturnian lightning strike. The best case scenario would require $10^{15}$ of such Saturnial lightning strikes to produce an observation signal in radio frequency of $\approx 4$mJy during a observation time $\approx 40$mins. This translates in a lightning flash density of 114 flashes km$^{-2}$ h$^{-1}$. The parameter study in 
\citep{2016MNRAS.461.1222H} provides insight into the challenges involved. Such Saturnian lightning on the mini-Neptune HAT-P-11b would produce 2-8 times more radio power than terrestrial lightning but time scales involved are rather short. Chemical tracers are therefore another option to trace the effect of lightning in planetary atmospheres. HCN is one candidate tracer because once created by the ion-neutral chemistry associated with lightning, it can be mixed up into atmospheric layers that may be observationally accessible. HCN would survive for 2 - 3 years in a planetary atmosphere with a moderate effective vertical mixing.

\section{The presence of \ce{H_3^+} and \ce{H_3O^+} on brown dwarfs}\label{s:ALSR}

It is likely that brown dwarfs have aurorae because  brown dwarfs possess (strong) confining (magnetic) field, free charges in form of electrons and their collisional partners in form of the atmospheric gas. Cyclotron maser emission has been detected from several brown dwarfs \citep{Hallinan2008}, which are produced by electron beams \citep{Schneider1959}, plausibly from energetic electrons transported through the atmosphere along magnetic field lines \citep{Nichols2012}, colliding with an ionizing the neutral species they encounter, just as auroral electrons ionize atmospheric species on Jupiter and Saturn \citep{Grodent2015}. Saturn and Jupiter receive their energetic electrons causing their auroral emission from their moons, Io and Europa, and Mercury receives its energetic electrons as part of the solar wind. Brown dwarfs can not be argued to possess moons easily as they form like stars by gravitational collapse, and not like planets by collisions of boulders of different sizes in a protoplanetary disk. The energetic electrons required for an aurora must therefore come from the brown dwarf's atmosphere. Brown dwarfs can be expected to form an ionosphere in their outermost regions as result of external irradiation as summarized in Sect.~\ref{s:GI}\ref{ss:environ}. Optical auroral emission has been observed on LSR-J1835 \citep{2015Natur.523..568H}, an ultracool star hugging the stellar vs. substellar boundary. We use LSR-J1835 as a representative case in what follows, modelling its atmospheric chemistry in the presence of UV irradiation from the interstellar medium, galactic cosmic rays, and ionization via auroral electrons, each of which contribute to the formation of ions, principally \ce{H_3^+} and \ce{H_3O^+}.

\subsection{The STAND2019 Chemical Network and ARGO high-energy chemistry model}

Ion abundances deep in the atmospheres of brown dwarfs is close to chemical equilibrium \citep{Lavvas2014,Rimmer2014,Rimmer2016}. If there were no UV photons, no energetic particles, no intrinsic disequilibrium processes active in the brown dwarf atmosphere, the results will not depart far from equilibrium except in the exosphere. Three body recombination deep in the atmosphere, dissociative recombination, and ion-neutral reactions, are fast ($\sim 10^{-9}$ cm$^3$ s$^{-1}$ rate constants for most ion-neutral two-body reactions), and effectively barrier-less. Ions shouldn't be quenched by vertical mixing.

The production of ions in the upper atmosphere of brown dwarfs is dominated by disequilibrium processes, such as photochemistry and energetic particle chemistry. To predict the effect of these processes on mixing ratios throughout the brown dwarf atmosphere, we solve the 1D Diffusion-Energetic Chemistry Equation:
\begin{equation}
\dfrac{d n_i}{dt} = P_i - L_i - \dfrac{\partial \Phi_i}{dz},
\end{equation}
where $n_i$ [cm$^{-3}$] is the number density of species $i$, $t$ [s] is time, $P_i$  is the rate of production of that species, and $L_i$ [cm$^{-3}$ s$^{-1}$] is the loss rate. The term $\partial \Phi_i/dz$ [cm$^{-3}$] s$^{-1}$] describes the vertical diffusion. We employ, ARGO, a Lagrangian chemical kinetics model \citep{Rimmer2016}, which is solved in the frame following a parcel of gas through the atmosphere, where:
\begin{equation}
\dfrac{\partial \Phi_i}{dz} \rightarrow \dfrac{\partial \Phi_i}{dz} - \dfrac{\partial \Phi_0}{dz}.
\label{eqn:Lagrange}
\end{equation}
Below the homopause, $\partial \Phi_i / dz \approx \partial \Phi_0 / dz$, and so there:
\begin{equation}
\dfrac{d n_i}{dt} \approx P_i - L_i .
\end{equation}
Eq. (\ref{eqn:Lagrange}) is folded into the production and loss terms, which change with time dependent on the parcel's location in the atmosphere and the local temperature, pressure, UV and particle fluxes, as described by \citet{Rimmer2016}. Once the parcel returns to the base of the atmosphere, transport of ultraviolet photons and energetic particles is calculated, and depth-dependent ionization rates are determined. The chemistry is solved again using these rates, after which the photon and energetic particle transport is calculated again. These two calculations are iterated until the code converges on a solution. 

ARGO employs the STAND2019 chemical network, which arises from the STAND2016 network \citep{Rimmer2016}, with updates to rate coefficients and the modification and addition of several reactions to better represent experimental results and observations of the Earth's atmosphere \citep{Rimmer2019}. STAND2019 includes over 5000 reactions incorporating H/C/N/O, complete for two carbon species, two nitrogen species, and three oxygen species, and valid for temperatures between 200 K and 30000 K. It has been benchmarked against the modern Earth, Jupiter \citep{Rimmer2016}, and hot Jupiter models \citep{Tsai2017,Hobbs2019}, and has been applied to ultra-hot Jupiters \citep{Kitzmann2018,Hoeijmakers2018}, early Earth atmospheres \citep{Rimmer2019b}, lightning in exoplanet atmospheres \citep{2017pre8.conf..345H,Ardaseva2017}, and terrestrial magma chemistry \citep{Rimmer2019c}.

\begin{figure}
\centering
\begin{subfigure}[b]{0.49\textwidth}
\caption{Temperature\label{fig:bd-temp}}
\includegraphics[width=\textwidth]{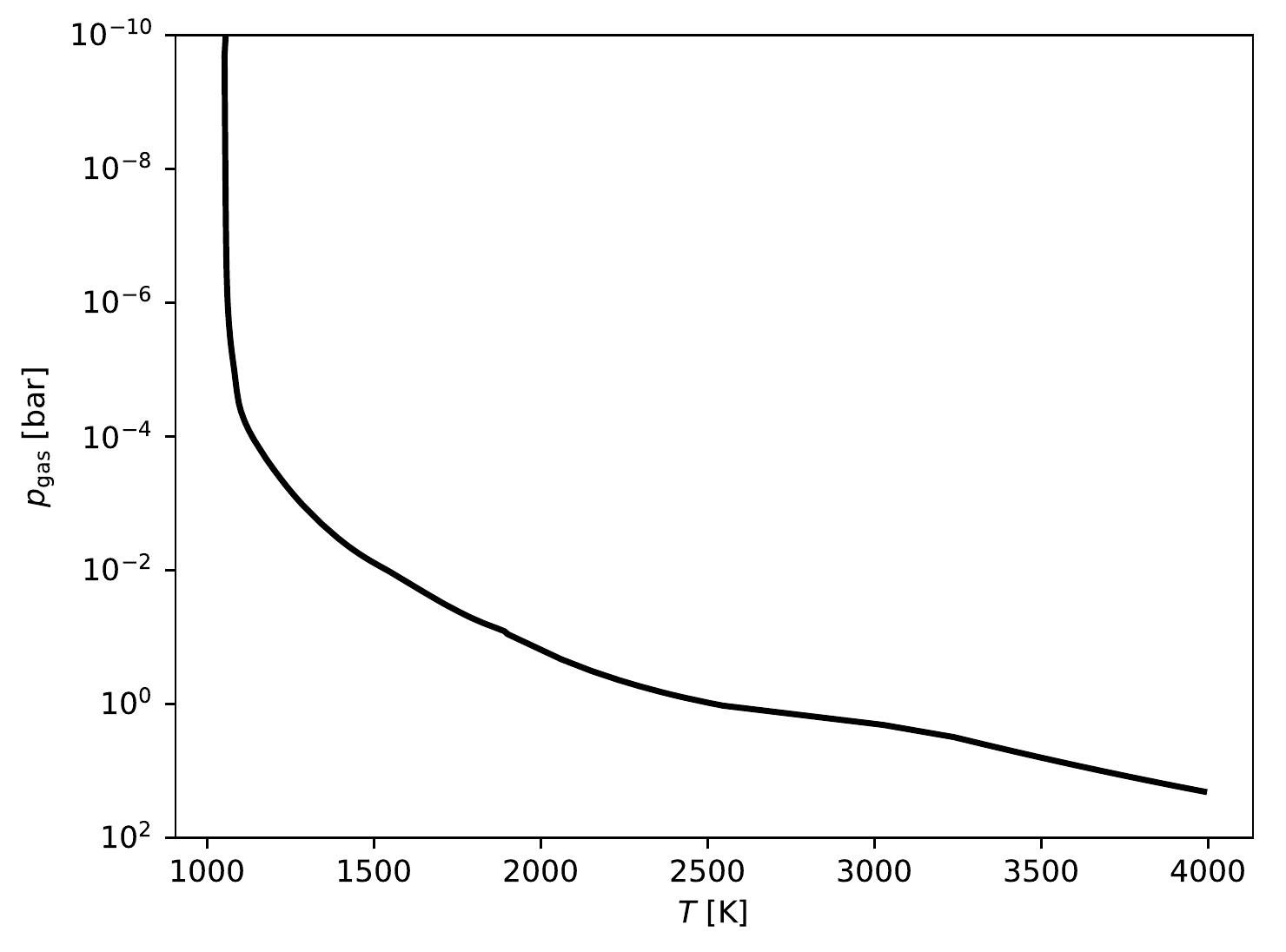}
\end{subfigure}
\begin{subfigure}[b]{0.49\textwidth}
\caption{Eddy Diffusion\label{fig:bd-kzz}}
\includegraphics[width=\textwidth]{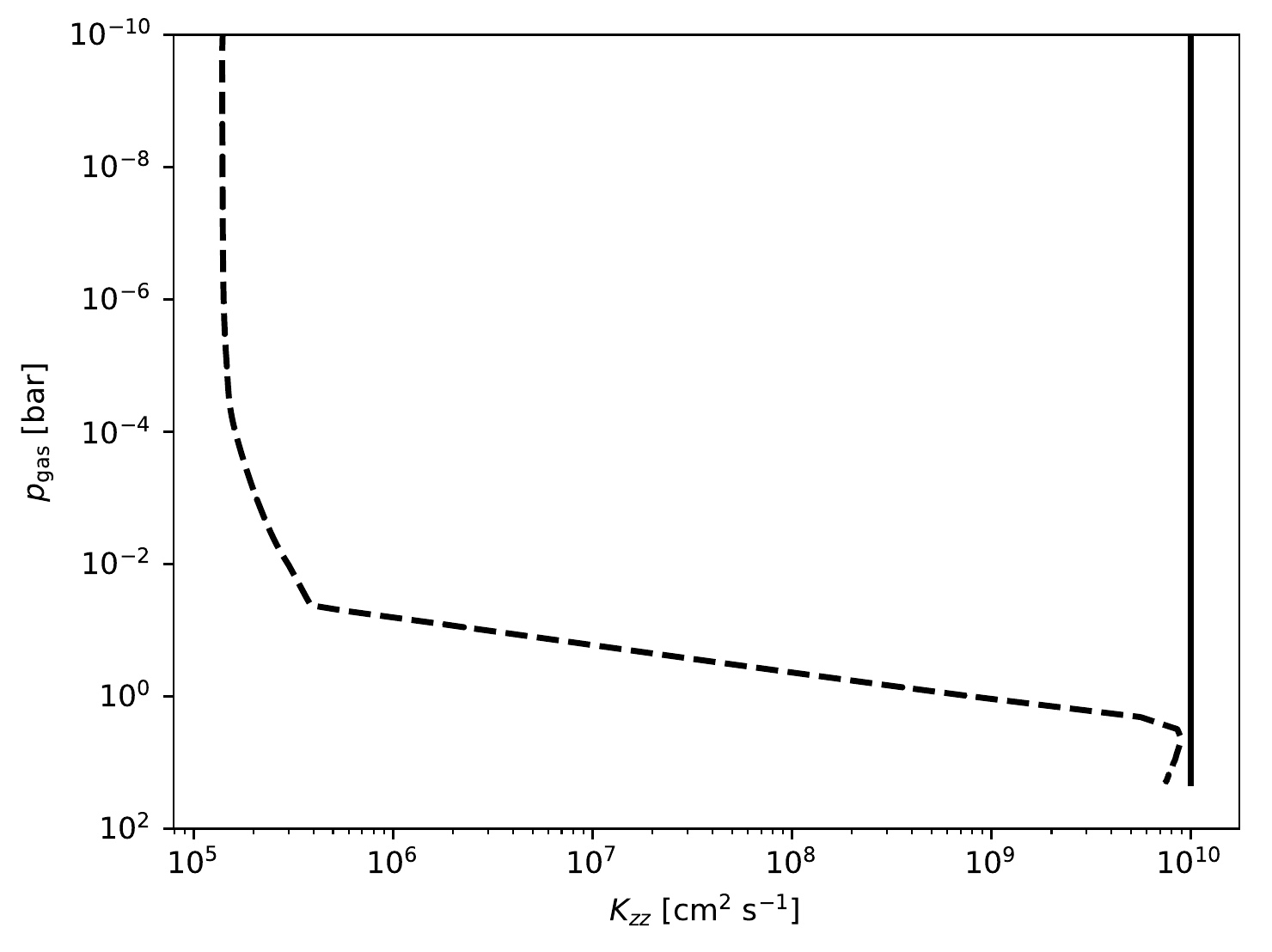}
\end{subfigure}
\caption{Temperature [K] (Fig. \ref{fig:bd-temp}), and $K_{zz}$ [cm$^2$ s$^{-1}$] (Fig. \ref{fig:bd-kzz}), as a function of pressure for a PHOENIX model M 8.5 dwarf, with effective temperature of $T_{\rm eff} =  2600$ K, surface gravity of $\log g = 5$ and the mean molecular mass of 2.33 amu.\label{fig:bd-profile}}
\end{figure}

We apply this chemical network to a PHOENIX model M 8.5 dwarf atmosphere \citep{2000ApJ...539..366A,2012RSPTA.370.2765A}, used to represent LSR-J1835. This model atmosphere has an effective temperature of $T_{\rm eff} =  2600$ K, surface gravity of $\log g = 5$ and the mean molecular mass of 2.33 amu. ARGO also needs a timescale for vertical mixing. We can estimate this timescale, $t_z$ [s] and Eddy diffusion coefficient, $K_{zz}$ [cm$^2$ s$^{-1}$] from the convective velocity, $v_{\rm conv}$ [cm/s], one of the PHOENIX model outputs, using Eq's (6) and (7) from \citet{2015A&A...580A..12L}:
\begin{align}
t_z &= \dfrac{H_0}{v_{\rm conv}}, \label{eqn:timescale}\\
K_{zz} &= v_{\rm conv} H_0,
\end{align}
where $H_0$ [cm] is the atmospheric scale height. Eddy diffusion is driven not simply by bulk convection but by the turbulent motion within convective cells, and may be much higher. For our calculations, we use a constant $K_{zz}$ = $10^{10}$ cm$^2$ s$^{-1}$, but we will use both the estimate from Eq. (\ref{eqn:timescale}) and this constant value when comparing with the chemical timescales below. The temperature profile and Eddy Diffusion coefficients are shown in Figure \ref{fig:bd-profile}. 

\subsection{Photochemical generation of \ce{H_3^+} on brown dwarfs}\label{ss:h3+}

Photoionization of \ce{H_2} in a gas giant atmosphere, whether within \citep[e.g.][]{Gross1964, Atreya1976}, or outside our Solar System \citep{Miller2000,Miller2013}, will readily lead to the formation of \ce{H_3^+} via the reactions:
\begin{align}
\ce{H_2} + h\nu &\rightarrow \ce{H_2^+} + e^-, \label{eqn:ionize}\\
\ce{H_2^+} + \ce{H_2} &\rightarrow \ce{H_3^+} + \ce{H}.
\label{eqn:ionize2}
\end{align}
If \ce{H_2} is the most probable second body, then virtually all of the \ce{H_2^+} will react with \ce{H_2} to form \ce{H_3^+}, and \ce{H_3^+} will be destroyed by either dissociative recombination with its electron:
\begin{align}
\ce{H_3^+} + e^- &\rightarrow \ce{H_2} + \ce{H},
\label{eq:direc}\\
 &\rightarrow 3\ce{H}.
\end{align}
The steady-state concentration of \ce{H_2} will be dependent on actinic flux of UV photons (the number of photons per cm$^2$ per second per \AA\ integrated over a unit sphere), multiplied by the cross-section for Reaction (\ref{eqn:ionize}). This cross-section is very large ($\sim 10^{-17}$ cm$^2$) for energies above near the ionization energy for \ce{H_2} ($\gtrsim 15.9$ eV). This means that Reaction (\ref{eqn:ionize}) is very efficient, even for a relatively low actinic flux, but also that \ce{H_2} will self-shield over a relatively short distance. 

\begin{figure}
\includegraphics[width=\textwidth]{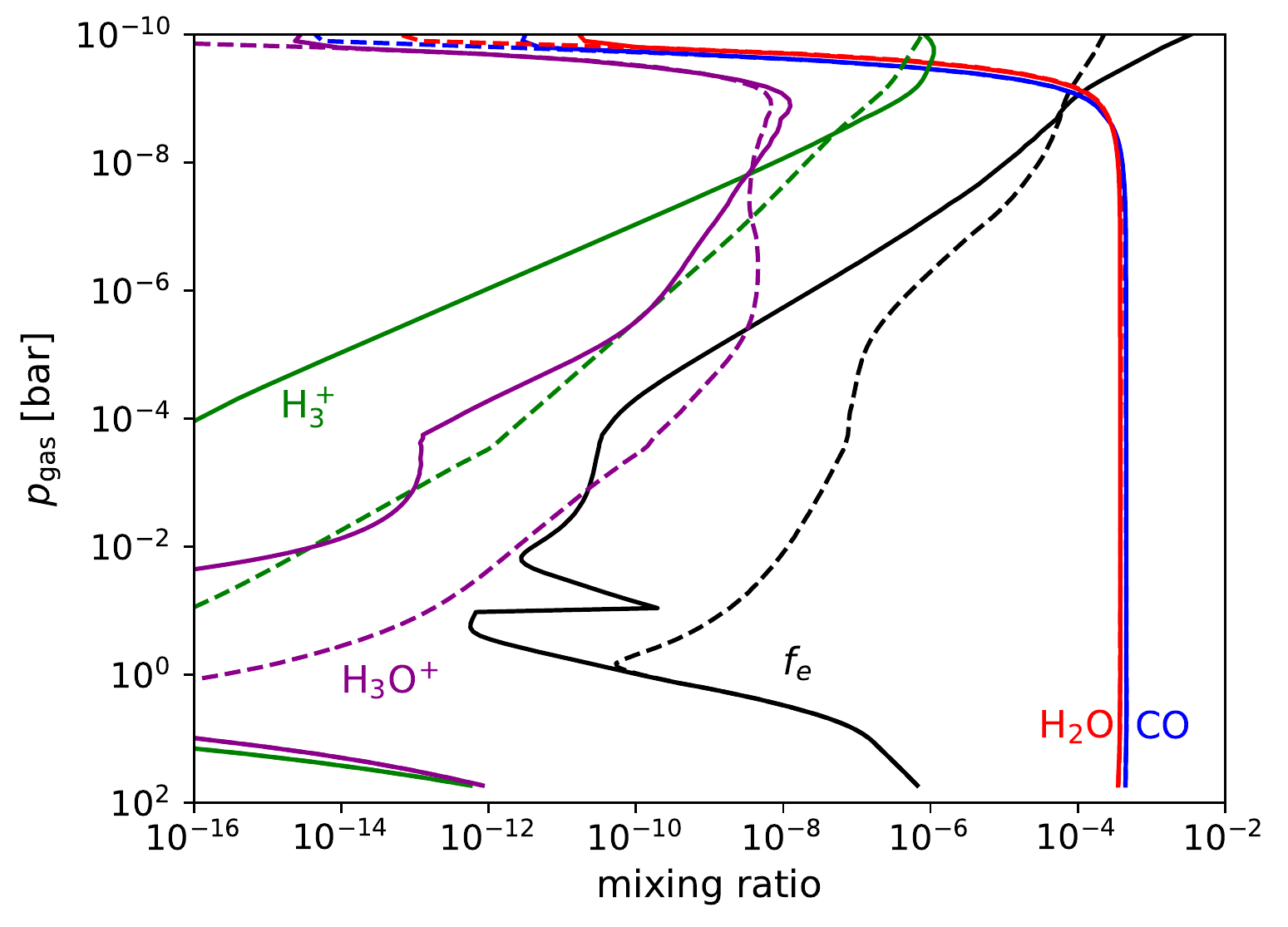}
\caption{Mixing ratios vs pressure [$p_{\rm gas}$, bar] for the ions \ce{H_3^+} and \ce{H_3O^+}, as well as the species involved in their destruction ($e^-$, \ce{CO} and \ce{H_2O}). Solid lines represent the results when accounting for photochemistry and cosmic ray chemistry, and dashed lines represent the results when including an auroral electron beam, as described in Section \ref{sec:aurora}.\label{fig:aurora-mix}}
\end{figure}

\begin{figure}
\includegraphics[width=\textwidth]{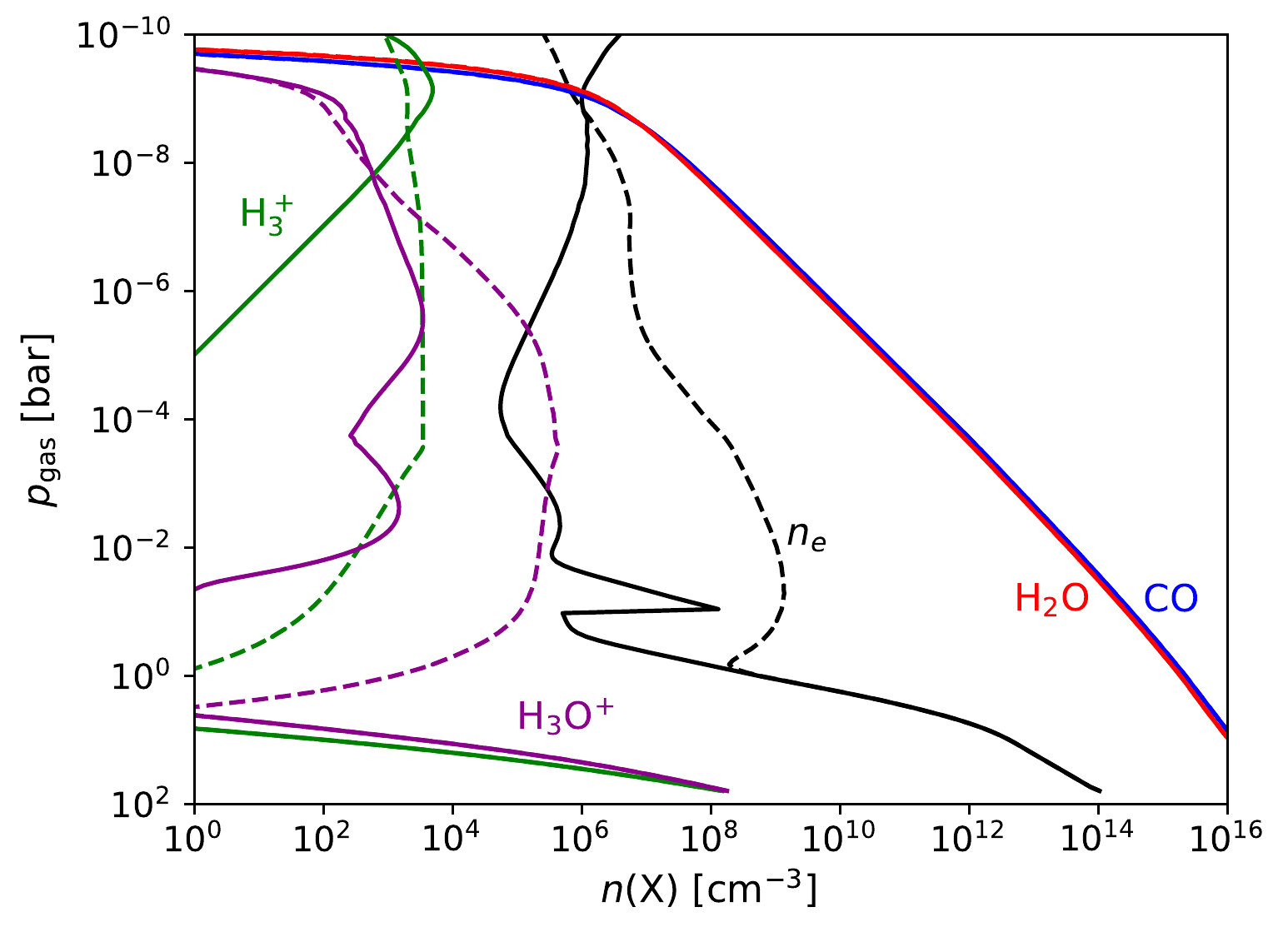}
\caption{Gas-phase number density ($n$ [cm$^-3$]) vs pressure ($p_{\rm gas}$ [bar]) for the ions \ce{H_3^+} and \ce{H_3O^+}, as well as the species involved in their destruction ($e^-$, \ce{CO} and \ce{H_2O}). Solid lines represent the results when accounting for photochemistry and cosmic ray chemistry, and dashed lines represent the results when including an auroral electron beam, as described in Section~\ref{s:ALSR} \ref{sec:aurora}.}
\label{fig:aurora-density}
\end{figure}

Figures \ref{fig:aurora-mix} and \ref{fig:aurora-density} show that some \ce{H_3^+} is generated within the thermosphere and exosphere of a brown dwarf from interstellar ultraviolet irradiation alone. Interstellar cosmic rays also contribute to the generation of \ce{H_3^+} in the upper atmospheres of free-floating ultracool stars. 

\subsection{Mechanism for Generating \ce{H_3^+} by Electron Beam}

Strong electron beams have been inferred in the atmospheres of some ultracool stars, such as LSR-J1835 \citep{2015Natur.523..568H}. The mechanism for generating \ce{H_3^+} from an electron beam is identical to the mechanism above, except for Reaction (\ref{eqn:ionize}), which is instead \citep{Miller2000}:
\begin{equation}
\ce{H_2} + e^{-,*} \rightarrow \ce{H_2^+} + e^{-,*} + e^-,
\end{equation}
where $e^{-,*}$ represents a high-energy electron. For the ionization cross-section, $\sigma_i(E)$, we use the values given by \citet{Padovani2009,Rimmer2013}. This chemical mechanism for generating \ce{H_3^+} is similar to the mechanism discussed for hot Jupiters by \citet{Chadney2016}.

Although the chemistry is very similar between electron-induced ionization and photoionization, the physical process is different, and so the part of the atmosphere effected is also different. In order to model the effect of electron-induced ionization in a Brown Dwarf atmosphere, we need to consider the incident flux of electrons. We will use the aurora detected on the M 8.5 dwarf LSR-J1835 as a representative auroral electron beam impinging on brown dwarf atmospheres. As we discussed above, the intensity of the cyclotron maser emission (CME) from LSR-J1835 is $10^5$ times that of Jupiter. This can be explained either by shifting the electron energy up, or by increasing the number of electrons by $10^5$. Shifting the electron energy too far will result in electron synchrotron emission, and will thus remove the CME. More intense CME's are best explained by a greater number of keV electrons \citep{Vorgul2011}. On this basis, and the observed CME intensity from LSR-J1835, we simply take the flux for the auroral electron beam for Jupiter \citep{Gerard1982}, and multiply it by $10^5$. For the Jovian auroral electron beam, we use the form of \citet{Gerard1982}, where $E$ [eV] is the electron energy, $j_0 = 1.25 \times 10^{15}$ electrons cm$^{-2}$ s$^{-1}$ eV$^{-1}$, and $E_0$ [eV] is the characteristic electron energy. We use 5 keV for the characteristic electron energy, as inferred by \citet{Gerard2009}:
\begin{equation*}
j(E) = \dfrac{j_0}{10^5} \Bigg(\dfrac{E}{E_0}\Bigg) e^{-E/E_0},
\end{equation*}
and multiply it by $10^5$, to yield:
\begin{equation}
j(E) = j_0 \Bigg(\dfrac{E}{E_0}\Bigg) e^{-E/E_0}.
\end{equation}
This assumes that the energy of the electron beam scales with the overall auroral energy, but since the auroral electron energy is estimated from the cyclotron maser emission observed on LSR-J1835, and the intensity of this emission depends on the number of electrons, this seems to be a reasonable approximation. We could use other methods we could use to adapt a Jovian auroral electron beam to LSR-J1835, such as shifting the energy of the electrons instead of increasing their number, but we could only adopt this method to a limited extent, because cyclotron maser emission require nonrelativistic electrons, and so limits us to electrons of energy $\lesssim 500$ keV. As brown dwarfs are exposed to harsher external radiation fields leading to highly ionised upper atmospheres \citep{2018A&A...618A.107R}, it is most plausible that the difference in intensity of the aurora on LSR-J1835, compared to Jupiter, is due to a greater number of electrons. 

In order to find how the energy-dependent flux of the electron beam evolves as the electron beam penetrates from the top into the atmosphere, we follow the same approach of \citep{2012A&A...537A...7R}, where we apply a Monte Carlo model with 100,000 test electrons which are injected into the top of the atmosphere with initial energies, $E_i$ representative of the energy-dependent flux appropriate to the electron bream. At each step through the atmosphere, $dz$, each electron is assigned a random value distributed uniformly between [0,1], and this value is compared to the probability, $P(E_i)$ that an electron of energy $E_i$ [eV] would experience an inelastic collision:
\begin{equation}
P(E_i) = n_{\rm gas}\sigma(E_i) \, dz,
\end{equation}
where $n_{gas}$ [cm$^{-3}$] is the gas-phase density, and $\sigma(E_i)$ [cm$^2$] is the total cross-section for an interaction. If the random number is less than $P(E_i)$, an interaction occurs, and a second random number, uniformly distributed between [0,1], is assigned to the electron, and this number is compared to the normalized cross-section to dissociate ($\sigma_d$), excite ($\sigma_{JJ'}$) or ionize ($\sigma_i$) the electron, where:
\begin{equation}
\Sigma_{J'}\sigma_{JJ'}(E) + \sigma_d(E) + \sigma_i(E) = \sigma(E).
\end{equation}
Each of these interactions has a characteristic energy loss, $W$ [eV], which is subtracted from the energy of the electron. The electrons are binned by energy after $dz$, and this is the updated spectrum of the electron beam. This is the same as the Monte Carlo model for cosmic ray energy loss presented by \citet{Rimmer2012} and \citet{Rimmer2013} but for electrons instead of cosmic ray protons.

Auroral electrons penetrate much more deeply into brown dwarf atmospheres than UV photons, and as a result the \ce{H_3^+} profile extends much further into the brown dwarf atmosphere, as can be seen in Figures \ref{fig:aurora-mix} and \ref{fig:aurora-density}. The \ce{H_3^+} concentration drops off precipitously approaching 1 bar, and this indicates the attenuation of the electron beam.

\subsection{Destruction of \ce{H_3^+} and the formation of \ce{H_3O^+} in brown dwarf atmospheres}
\label{sec:aurora}

In the upper atmosphere, \ce{H_3^+} is primarily destroyed by recombination with its electron. Because \ce{H_3^+} reacts rapidly with common constituents in Brown Dwarf atmospheres, \ce{CO} and \ce{H_2O}, its chemical lifetime overall is short and density-dependent. The relevant reactions are:
\begin{align}
\ce{H_3^+} + e^- &\rightarrow {\rm Products}, & k_1 = 2.8 \times 10^{-8} \; {\rm cm^3 \, s^{-1}} \, \Bigg(\dfrac{T}{300 \, {\rm K}}\Bigg)^{\! -0.52}; \\
\ce{H_3^+} + \ce{CO} &\rightarrow \ce{HCO^+} + \ce{H_2}, & k_2 = 1.4 \times 10^{-9} \; {\rm cm^3 \, s^{-1}}; \\
\ce{H_3^+} + \ce{H_2O} &\rightarrow \ce{H_3O^+} + \ce{H_2} & k_3 = 4.3 \times 10^{-9} \; {\rm cm^3 \, s^{-1}}. \label{eqn:hydronium-prod}
\end{align}
The ion \ce{HCO^+} reacts very quickly with other atmospheric constituents and its abundance is very low, with mixing ratios $< 10^{-16}$ throughout. Hydronium (\ce{H_3O^+}), on the other hand, is more stable, and becomes the dominant hydrogen-bearing ion in the Brown Dwarf's lower atmosphere (see Fig's \ref{fig:aurora-mix}, \ref{fig:aurora-density}). The \ce{H_3O^+} is destroyed by electrons and ammonia, with the reactions:
\begin{align}
\ce{H_3O^+} + e^- &\rightarrow {\rm Products}, & k_4 = 4.3 \times 10^{-7} \; {\rm cm^3 \, s^{-1}} \, \Bigg(\dfrac{T}{300 \, {\rm K}}\Bigg)^{\! -0.5}; \\
\ce{H_3O^+} + \ce{NH_3} &\rightarrow \ce{NH_4^+} + \ce{H_2O}, & k_5 = 2.5 \times 10^{-9} \; {\rm cm^3 \, s^{-1}}.
\end{align}
Since these are the dominant destruction pathways for \ce{H_3^+} and \ce{H_3O^+} , and the products do not cycle back to reform \ce{H_3^+} and \ce{H_3O^+}, the above rate constants can be used directly to estimate the chemical lifetimes of these cations.

\begin{figure}
\includegraphics[width=\textwidth]{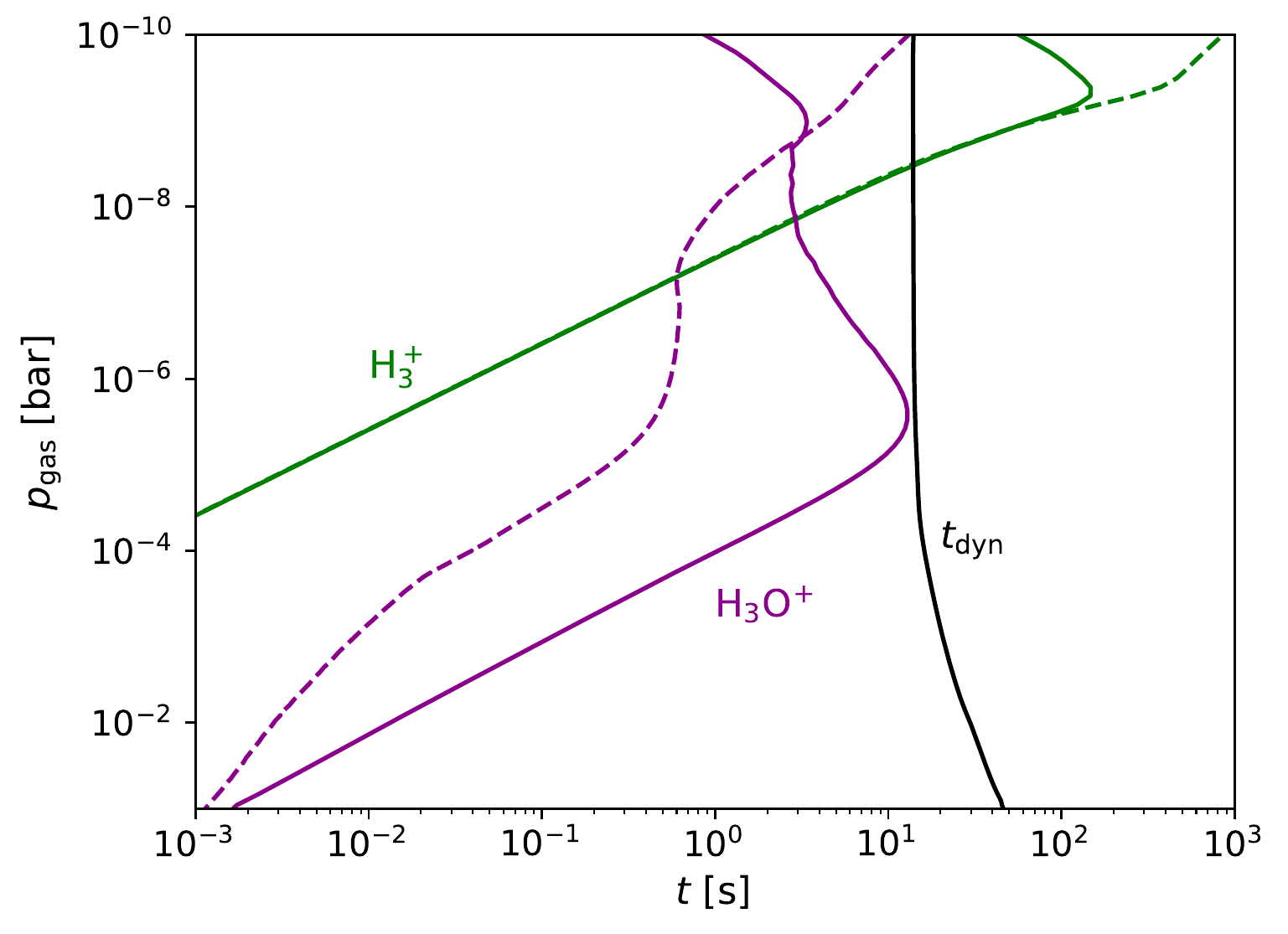}
\caption{Chemical timescales for \ce{H_3^+} and \ce{H_3O^+}, ranging from $\lesssim 1$ ms where $p_{\rm gas} > 10^{-1}$ bar to 10-100 seconds where $p_{\rm gas} < 10^{-4}$ bar. Solid lines represent the results when accounting for photochemistry and cosmic ray chemistry, and dashed lines represent the results when including an auroral electron beam, as described in Section \ref{sec:aurora}. The black solid line represents the dynamical timescale, $t_{\rm dyn}$ [s], corresponding to a contstant $K_{zz} = 10^{10}$ cm$^{2}$ s$^{-1}$.) \label{fig:aurora-timescale}.}
\end{figure}

For \ce{H_3^+}, the chemical lifetime is:
\begin{equation}
\tau_{\rm chem}(\ce{H_3^+}) = \dfrac{[\ce{H_3^+}]}{d[\ce{H_3^+}]/dt} = \dfrac{1}{k_1[e^{-1}] + k_2[\ce{CO}] + k_3[\ce{H_2O}]},
\end{equation}
and for \ce{H_3O^+}:
\begin{equation}
\tau_{\rm chem}(\ce{H_3O^+}) = \dfrac{[\ce{H_3O^+}]}{d[\ce{H_3O^+}]/dt} = \dfrac{1}{k_4[e^{-1}] + k_5[\ce{NH_3}]}.
\end{equation}
The chemical timescales for \ce{H_3^+} and \ce{H_3O^+} are shown in Fig. \ref{fig:aurora-timescale}. These timescales can be compared with dynamic timescales, $t_{\rm dyn}$, estimated using vertical or horizontal mixing velocities. When $t_{\rm dyn} > t_{\rm chem}$, the chemistry is driven out of equilibrium by photodissociation and photoionization, but is not much influenced by the dynamics of the fluid motion in the atmosphere. When $t_{\rm dyn} > t_{\rm chem}$, on the other hand, the dynamics dominates, and the species can be transported into regions with concentrations far from equilibrium. If we compare the chemical timescale to the dynamical timescale calculated using Eq. (\ref{eqn:timescale}), we find the dynamical timescale is orders of magnitude larger than the chemical timescale throughout the brown dwarf atmosphere, and the ion chemistry is never quenched. On the other hand, if we compare the chemical timescale to the timescale derived from our constant $K_{zz} = 10^{10}$ cm$^2$ s$^{-1}$ (as in Fig. \ref{fig:aurora-timescale}), we find that the chemical timescale exceeds the dynamical timescale in the upper atmosphere, and so the ion chemistry may be quenched in the upper atmosphere with efficient vertical mixing.

\enlargethispage{20pt}

\section{Conclusion}

The atmospheres of brown dwarfs and hot Jupiters are witness to physical and chemical processes in  regimes that can be very different from anything in our Solar System. Atmospheric temperatures range between $<300\,\ldots\,\sim 3500$K for irradiated planets, and the contrast between day and night can be extreme, resulting in planet-scale equatorial jets that respond to the steep temperature gradients. Cloud particles form from seed particles made of, for example, titanium dioxid and silicon monoxid, that grow a mantle containing a mix of  Mg/Si/Fe/Al/Ti/C/$\ldots$/O minerals, fall through the atmosphere like rain, until they are dissolved deep in the atmosphere by temperatures high enough to sublimate iron. For all these extremes, hot Jupiters and brown dwarfs share some surprising similarities with the gas giants in our Solar System. The particles (aerosols)  that make up their clouds can become ionized, and the electrification and subsequent advection of cloud particles generates charge separation over atmospheric scales, and may lead to lightning. Aurorae on brown dwarfs, though $10^5$ times more intense, and although probably not generated by a moon or rings, are explained by the same fundamental mechanisms that generate aurorae on Jupiter and Saturn. The source of the electrons, the location of the electron-gas interaction, and the relative abundances of the gas species are different but the immediate chemical products of the electron-gas interactions, \ce{H_2^+} and \ce{H_3^+}, are the same within these highly diverse, but hydrogen-dominated atmospheres.

We find that the chemistry that leads to \ce{H_3^+} generation on hot Jupiters and brown dwarfs is the same as the chemistry on Jupiter and Saturn. The gas giants in our solar system are very cold, so  that water freezes out much deeper in their atmospheres than for extrasolar hot Jupiters and warm brown dwarfs. Therefore,  less H$_2$O and CO remains in the gas phase compared to hot Jupiters, L-type brown dwarfs and very late M-type dwarfs. Consequently, when present, the water vapor and carbon monoxide react readily with \ce{H_3^+} to form \ce{HCO^+} and hydronium (\ce{H_3O^+}), respectively. Interstellar ultraviolet radiation and galactic cosmic rays generate \ce{H_3^+} at microbar to nanobar pressures, and \ce{H_3O^+} at millibar to microbar pressures. The energetic electrons that would drive aurorae on brown dwarfs ionize hydrogen much deeper in their atmosphere, generating $10^6$ cm$^{-3}$ densities of \ce{H_3O^+} at 1 bar. 

Producing \ce{H_3^+} requires molecular hydrogen, and this may be one of the reasons \ce{H_3^+} has proved difficult to detect in hot Jupiter atmospheres. The irradiation would ionize \ce{H_2}, but also dissociates \ce{H_2}. The 10000 K thermospheric temperatures expected for Hot Jupiters \citep{Koskinen2010}, along with the intense radiation, means that the most prevalent neutral species at these heights is atomic hydrogen, and the most abundant ion is \ce{H^+}. Since $f_e/f(H_3^+) > 1$ in the upper atmospheres of Hot Jupiters, it will be difficult to sustain much \ce{H_3^+}, as has already been shown by \citet{Chadney2016}. In the deeper, radiation-sheltered atmosphere, \ce{H_3^+} will again be destroyed by collisions with H$_2$O and CO similar to brown dwarfs.
Brown dwarfs, on the other hand, may have much cooler upper atmospheres, amenable to the stability of \ce{H_2}, and therefore the efficient production of \ce{H_3^+}, either via galactic cosmic rays and interstellar UV irradiation, or by collisional ionization with the high energy electrons that would generate brown dwarf aurorae. The lifetime of \ce{H_3^+} is very short in the deeper atmosphere, primarily because it reacts quickly with CO and \ce{H_2O}. The reaction of \ce{H_3^+} with \ce{H_2O} results in \ce{H_3O^+}. We propose searching for \ce{H_3^+} and \ce{H_3O^+} in free-floating brown dwarfs. For robust identification, absorption cross-sections valid for the extreme temperatures of these sub-stellar objects are needed. The ExoMol team already provides the needed \ce{H_3^+} line-lists \citep{Mizus2017}, but updated line-lists for \ce{H_3O^+} will be needed if this molecule is to be positively identified in a brown dwarf atmosphere.



\aucontribute{Ch.H. lead the paper, carried out the cloud and equilibrium gas phase calculations. P.B.R. performed the chemical kinetics calculations, made Fig's 3-6. Both authors jointly wrote the paper.}

\competing{The author have no competing interests.}

\funding{P.B.R. thanks the Simons Foundation for funding (SCOL awards 599634). Ch.H. and P.B.R. acknowledge funding from the European commission under which part of this research was conducted. We highlight financial support of the European Union under the FP7 by an ERC starting grant number 257431.}

\ack{We thank Jonathan Tennyson, Steve Miller and Benjamin McCall for organising an engaging  Royal Society discussion meeting {\it  Advances in Hydrogen Molecular Ions: H$_3^+$, H$_5^+$ and beyond}. We thank Sergey Yurchenko for useful discussions about high-temperature ion cross-sections. We thank Piere Gourbin for providing Fig.~1 and V. Parmentier for providing the cloud-free 3D GCM results for WASP-18b.}


\bibliographystyle{aa}
\bibliography{bib.bib}
\end{document}